\documentstyle[12pt,epsf]{article}
\title{\Large
$\sigma$ Exchange in the $NN$ Interaction within the Chiral Unitary Approach
}

\author{
E. Oset$^{1,2}$, H. Toki$^1$, M. Mizobe$^1$ and T. T. Takahashi$^1$
\bigskip\\
{\normalsize\it $^1$Research Center for Nuclear Physics(RCNP),Osaka University}\\
{\normalsize\it Mihogaoka 10-1, Ibaraki, Osaka 567-0047, Japan}\\
{\normalsize\it $^2$Departamento de F$\acute{\imath}$sica Te\'{o}rica and IFIC, Centro Mixto Universidad de}\\
{\normalsize\it Valencia-CSIC, 46100 Burjassot (Valencia), Spain}
}

\begin{document}
\maketitle

%\abst{%       %this abstract is neglected when [addenda] or [errata]
\begin{abstract}
We study the nucleon-nucleon interaction in the isoscalar-scalar channel using the chiral unitary approach. The $t$-matrix of the pion-pion scattering in this channel is summed up to all orders using the B-S equation. We find that the calculated results at long distances are close to those of the $\sigma$-exchange interaction.
In addition, there appears a shorter range repulsion in this channel.
%}
\end{abstract}

\section{Introduction}

The intermediate range attraction in the $NN$ interaction has been
traditionally described by the $\sigma$-exchange in the meson exchange picture.~\cite{R1} It has also been noticed~\cite{R2,R3,R4} that box diagrams with two-pion
exchange and intermediate $\Delta$ excitation lead to an intermediate range
attraction. A weakened $\sigma$ exchange together with
these box diagrams has also been used to describe the intermediate range $NN$ attraction.~\cite{R1}

With the success of chiral perturbation theory (${\chi}PT$) in the meson-meson and
meson-baryon sectors,~\cite{R5} attempts to extend these ideas to the $NN$
sector have been pursued.~\cite{R6,R7,R8}
In a recent paper,~\cite{R9} following the line of Ref. [8], the peripheral $NN$
partial waves are studied within a chiral scheme, using the meson-meson and
the meson-baryon interaction Lagrangians. With this input, together with more
conventional processes with $\Delta$ box diagrams, $\rho$ exchange and other meson exchange,
a good reproduction of the $NN$ data for $L>2$ partial waves is obtained.

A striking feature of Ref. [9] is that the exchange of two interacting pions
in the scalar-isoscalar channel (the $\sigma$ channel) leads to a
repulsion (although weak), instead of the commonly accepted attraction
from $\sigma$ exchange. Simultaneously, from the box diagrams with intermediate
$\Delta$, an attraction is obtained with the range and strength of the
standard $\sigma$ exchange of the boson exchange models. Two basic
approximations lead to these results. First, the $\pi$$\pi$ isoscalar 
interaction is used only
to lowest order in ${\chi}PT$. Second, no form factors are considered for the $\Delta$ box diagrams.

In the present work we wish to reconsider this idea and go further by
treating the exchange of two interacting pions in the isoscalar channel
in a nonperturbative way. The approach followed here for the $\pi$$\pi$
interaction produces a $\sigma$ pole in the complex plane in the physical
region $s>4m_{\pi}^2$. Then the analytical extrapolation of the
model is used in order to find the strength of the isoscalar exchange
interaction for the situation $s<0$, which one encounters in the $NN$
scattering problem.

The $\sigma$ meson has been rather problematic, with ups and
downs in the particle data tables, where it has been once again welcome.~\cite{R10}
Some analyses of the $\pi$$\pi$ data rely on the $\sigma$ pole
in the $\pi\pi$ {\it t} matrix.~\cite{R11,R12,R13} Theoretical models for the $\pi\pi$
interaction based on meson exchange~\cite{R14} find also a pole in the {\it t}
matrix for the $\sigma$ meson. Yet, of relevance to the present work is
the fact that the $\sigma$ pole is found in recent chiral nonpertubative
approaches which have developed independently.~\cite{R15,R16} In Ref. [15] the
inverse amplitude method using the lowest order and second order chiral
Lagrangians is formulated and it leads to a pole position for $\sigma$ at $440-i245$ MeV. In Ref. [16] the Bethe Salpeter equation, with
a cutoff in the loops fitted to the data of the scalar sector, is
shown to reproduce accurately the data in the scalar sector up to around
$\sqrt{s}=1.2$ GeV, using only the lowest order chiral Lagrangian as
input. In this case, $\sigma$ appears as a pole at $469-i203$ MeV [Ref. [16],erratum].
The approach of Ref. [16], using coupled channels, is able to reproduce
the $\sigma$ and $f_0$ (980) resonance in the scalar-isoscalar sector and
the $a_{0}$(980) resonance at $L=0$, $I=1$. The approach of Ref. [15] using
only the $\pi\pi$ channel could only generate $\sigma$ in the
scalar channel, but $\rho$ and $K^{*}$ in the vector channel were
accurately reproduced.

A generalization of the inverse amplitude method of Ref. [15], incorporating coupled channels as in Ref. [16] and using the lowest and
second order chiral Lagrangians, is given in Ref. [17], and there, all
meson-meson data up to $\sqrt{s}=1.2$ GeV are accurately reproduced. In the
latter work, the $\sigma$ pole appears at $442-i225$ MeV.

A different chiral approach is employed in Ref. [18], using the N/D method. This explicitly intoduces resonances on top of the lowest order chiral
Lagrangian. That approach, which includes contributions from the physical and 
unphysical cuts through dispersion relations, produces good results up to 
$\sqrt{s}=1.4$ GeV
and also leads to a $\sigma$ pole at $445-i221$ MeV.

As we can see, all these different schemes, which uses the lowest order chiral Lagrangian and the implementation of exact
unitarity, lead invariably to a $\sigma$ pole around $450-i225$ MeV, with
small fluctuations.

Our aim here is to follow these lines and use the unitary scheme to
generate the $\pi\pi$ scalar isoscalar amplitude in the unphysical (virtual)
region, and later its contribution to the $NN$ interaction.

\section{Lowest order contribution in isoscalar exchange in the $NN$ interaction}

\begin{figure}
\begin{center}
\epsfxsize=14cm
\centerline{\epsfbox{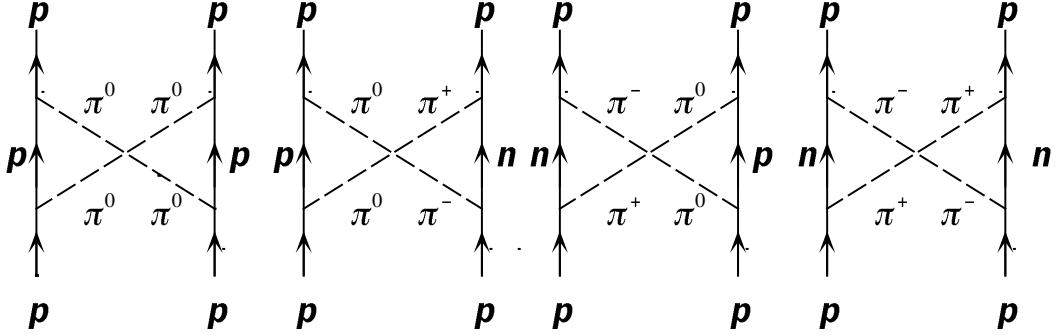}}
%\epsfile{file=figure1.eps,scale=0.8}
\caption{The lowest-order processes in the $\pi\pi$ interaction in the scalar isoscalar channel for pp$\to$pp}
\label{F1}
\end{center}
\end{figure}

We first investigate the lowest-order contribution to the $\pi\pi$
interaction in the scalar isoscalar channel, which is given by the
diagrams of Fig. \ref{F1} for $pp \to pp$. We leave out the box diagram contribution here and just refer the literature.~\cite{R1}

Considering the $\vec{\tau}$ $\vec{\phi}$ isospin dependence of
the $\pi NN$ coupling, one gets from the sum of the diagrams in Fig.\ref{F1} the
combination,
\begin{eqnarray}
t_{\pi^{0}\pi^{0}\to\pi^{0}\pi^{0}}+2t_{\pi^{0}\pi^{0}\to\pi^{+}\pi^{-}}
+2t_{\pi^{+}\pi^{-}\to\pi^{0}\pi^{0}}+4t_{\pi^{+}\pi^{-}\to\pi^{+}\pi^{-
}}.
\label{E1}
\end{eqnarray}
Taking into account the unitary normalization of the $\pi\pi$
states of Ref. [16], 
\begin{eqnarray}
|\pi\pi,I=0\rangle
=-\frac{1}{\sqrt{6}}|\pi^{0}\pi^{0}+\pi^{+}\pi^{-}+\pi^{-}\pi^{+}\rangle ,
\label{E2}
\end{eqnarray}
the combination of $t$-matrices in Eq. (\ref{E1}) corresponds, in terms of the isoscalar $\pi\pi$ amplitude, to 
\begin{eqnarray}
6t^{(I=0)}_{\pi\pi\to\pi\pi}.
\label{E3}
\end{eqnarray}
We should note here that all the pion lines in Fig. \ref{F1} are off shell,
which requires the use of the off shell $t$-matrix. From Ref. [16] this off shell
amplitude obtained from the lowest order meson-meson Lagrangian is given
by
\begin{eqnarray}
t^{(I=0,L)}_{\pi\pi\to\pi\pi}=-\frac{1}{9f^{2}}\left( 9s+\frac{15m_{\pi}
^{2}}{2}-3\sum_{i}p_{i}^{2}\right),
\label{E4}
\end{eqnarray}
where $f$ is the pion decay constant ($f=93$ MeV), $s$ is the $\pi\pi$
Mandelstam variable, and the $p_{i}$ are the momenta of the pion lines. We
define the on-shell value of the amplitude as that of Eq. (\ref{E4}) when 
$p_{i}^{2}=m_{\pi}^{2}$. This allows us to write in a convenient
way
\begin{eqnarray}
t^{(I=0,L)}_{\pi\pi\to\pi\pi}=t^{(I=0,L,OS)}_{\pi\pi\to\pi\pi}+\frac{1}{
3f^{2}}\sum_{i}(p_{i}^{2}-m_{\pi}^{2}),
\label{E5}
\end{eqnarray}
with the on shell value of the amplitude
\begin{eqnarray}
t^{(I=0,L,OS)}_{\pi\pi\to\pi\pi}=-\frac{1}{f^2}\left(s-\frac{m_{\pi}^2}{
2}\right).
\label{E6}
\end{eqnarray}

\begin{figure}[t]
\begin{center}
\epsfxsize=14cm
\centerline{\epsfbox{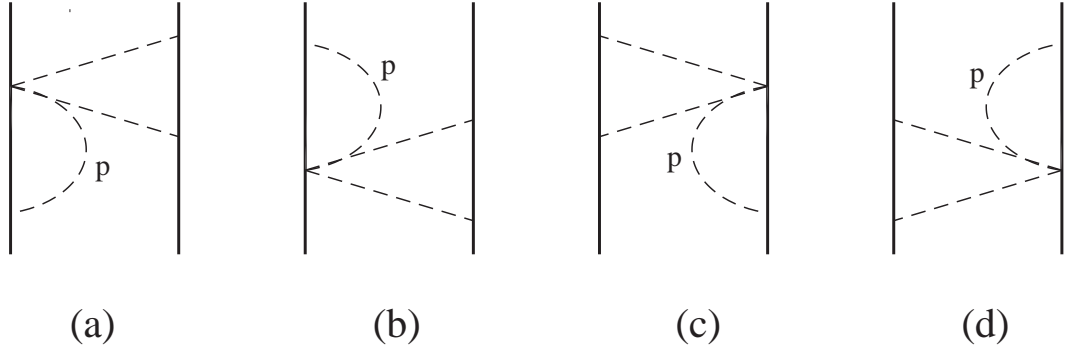}}
%\epsfile{file=figure2.eps,scale=1.2}
\caption{The processes with three meson vertex at a baryon line for $pp \to pp$.}
\label{F2}
\end{center}
\end{figure}

In what follows we prove that the off shell contribution of the
amplitude, i.e., the terms proportional to $p_{i}^{2}-m_{\pi}^2$, cancels
exactly with the diagrams of Fig. \ref{F2}, which appear at the same order of
the chiral counting.
The diagrams of Fig. \ref{F2} contain the vertices with one baryon line and
three mesons. The baryon-meson chiral Lagrangian relates this function to
that with one meson attached to the baryon line. This interaction
term is given by~\cite{R19}

\begin{eqnarray}
{\cal L}_{1}^{(B)}=\frac{D+F}{2}(\bar{p}\gamma^{\mu}\gamma_{5}u_{\mu}^{11}p+\bar{n}\gamma^{\mu}\gamma_{5}u_{\mu}^{22}n+\bar{n}\gamma^{\mu}\gamma_{5}u_{\mu}^{21}p+\bar{p}\gamma^{\mu}\gamma_{5}u_{\mu}^{12}n)
\label{E7}
\end{eqnarray}
where $p$ and $n$ represent the proton or neutron fields. In Eq. (\ref{E7}) $u_{\mu}$ is the
$SU(2)$ matrix given by
\begin{eqnarray}
u_{\mu}=-\frac{\sqrt{2}}{f}{\partial}_{\mu}\Phi+\frac{\sqrt{2}}{12f^3}({\partial}
_{\mu}\Phi\Phi^{2}-2\Phi{\partial}_{\mu}\Phi\Phi+\Phi^{2}{\partial}_{\mu}{\Phi}),
\label{E8}
\end{eqnarray}
and $\Phi$ is the $SU(2)$ matrix for the pions given by
\begin{eqnarray}
\Phi=\left(
\begin{array}{cc}
\frac{1}{\sqrt{2}}\pi^{0} & \pi^{+} \\
\pi^{-} & -\frac{1}{\sqrt{2}}\pi^{0}
\end{array}
\right).
\label{E9}
\end{eqnarray}

Equation(\ref{E7}), with the first term of $u_{\mu}$ in Eq. (\ref{E8}) gives rise to the
${\pi}NN$ coupling, which in the nonrelativistic reduction,
$\gamma^{\mu}\gamma_{5}\to\sigma^{k}\delta_{{\mu}k}$,
reads
\begin{eqnarray}
&&-it_{\pi^{i}NN}=C_{i}\frac{D+F}{2f}\vec{\sigma}\cdot\vec{p},\nonumber \\
&&C_{\pi^{+}}=C_{\pi^{-}}=\sqrt{2};\ C_{\pi^{0}pp}=1;\ C_{\pi^{0}nn}=-1
\label{E10}
\end{eqnarray}
for incoming pion with momentum $\vec{p}$.

\begin{figure}
\begin{center}
\epsfxsize=14cm
\centerline{\epsfbox{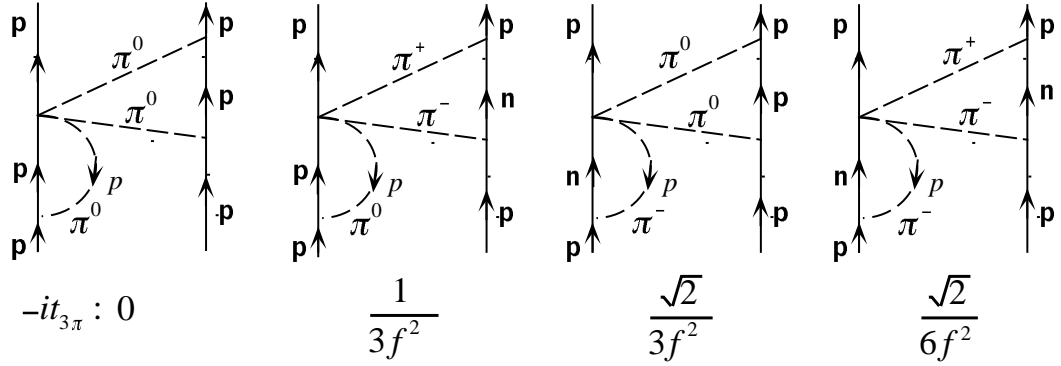}}
%\epsfile{file=figure3.eps,scale=0.8}
\caption{Diagrams involving the three pion vertex. The figure displays the coefficient which multiplies $\frac{D+F}{2f}\vec{\sigma}\vec{p}$ for each three pion vertex.}
\label{F3}
\end{center}
\end{figure}

The second term in the expansion of $u_{\mu}$ gives rise to the three
pion vertices and involves derivatives in the three pion fields.
However, given the fact that the ${\pi}NN$ vertex in the loops with
meson momentum $p$ of Fig. \ref{F2} contain the $p$-wave coupling
$\vec{\sigma}\vec{p}$, only the term which involves the derivative in the
pion with momentum $p$, which will give rise to another
$\vec{\sigma}\vec{p}$ vertex, will contribute in the loop integration.
In Fig. \ref{F3} we show the diagrams corresponding to the loop in Fig. \ref{F2}(a) for
the different isospin combinations of Fig. \ref{F1}, together with the
contribution of the three meson vertex to each diagram.

\begin{figure}
\begin{center}
\epsfxsize=6.6cm
\centerline{\epsfbox{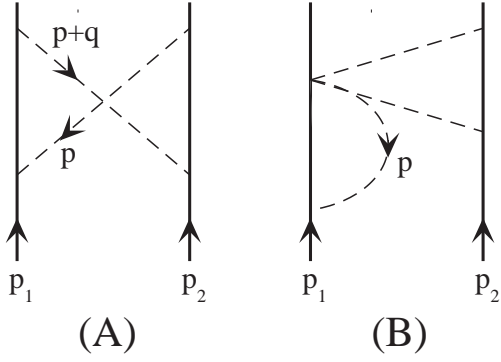}}
%\epsfile{file=figure4.eps,scale=1.2}
\caption{Diagrams where off-shell cancellations appear. Diagram (A) with the off shell part of the $\pi$ $\pi$ amplitude from the line $p+q$ cancels diagram (B).}
\label{F4}
\end{center}
\end{figure}

If one sums the contributions of all the diagrams in Fig. \ref{F3} including the isospin
weight of the ${\pi}NN$ vertices one obtains an equivalent three pion
vertex,
\begin{eqnarray}
-it_{3\pi}\equiv
\frac{1}{3f^2}\frac{D+F}{2f}\vec{\sigma}\cdot\vec{p}\left( 
2+2+2\right) =\frac{6}{3f^2}\frac{D+F}{2f}\vec{\sigma}\cdot\vec{p}.
\label{E11}
\end{eqnarray}
Now let $q$ be the momentum exchanged between the protons and let us
evaluate the diagrams of Fig. \ref{F1} with the off shell part of the meson-meson vertex and the diagrams of Fig. \ref{F3}. We present them in Fig. \ref{F4} with the appropriate momentum assignment.

The loops to the right in both diagrams of the figure are the same. Hence,
we concentrate only on the loop of the left. Let us pick up the term
$((p+q)^{2}-m_{\pi}^2)/3f^2$ of the meson-meson vertex, which provides
the off shell contribution of the left top meson in diagram (A). We then get

\begin{eqnarray}
V^{(A)}&=&i\left(\frac{D+F}{2f}\right)^{2}\int\frac{d^{4}p}{(2\pi)^4}
\vec{\sigma}\cdot(\vec{p}+\vec{q})\vec{\sigma}\cdot\vec{p}\ 
6\frac{1}{3f^2}[(p+q)^{2}-m_{\pi}^2]\frac{1}{p^2-m_{\pi}^2+i\epsilon}
\nonumber \\
&&\frac{1}{(p+q)^2-m_{\pi}^2+i\epsilon}\ \frac
{M}{E(\vec{p_1}+\vec{p})}\ \frac{1}{p_1^0
+p^0 -E(\vec{p_1}+\vec{p})+i\epsilon},
\label{E12}
\end{eqnarray}
with $M$ and $E$ the mass and on shell energy of the nucleon. Similarly, we 
obtain the contribution from the loop on the left in diagram (B), 
\begin{eqnarray}
V^{(B)}&=&-i\left(\frac{D+F}{2f}\right)^{2}\int\frac{d^{4}p}{(2\pi)^4}
\vec{\sigma}\cdot\vec{p}\ \vec{\sigma}\cdot\vec{p}\ 6\frac{1}{3f^2}\ 
\frac{1}{p^2 -m_{\pi^{2}}+i\epsilon}
\nonumber \\
&&\frac{M}{E(\vec{p_1}+\vec{p})}\ \frac{1}{p_{1}^0+p^0 -E
(\vec{p_1}+\vec{p})+i\epsilon}.
\label{E13}
\end{eqnarray}
In Eq. (\ref{E12}) one can see a cancellation of the off-shell part of 
the meson-meson vertex with the corresponding meson propagator. 
The rest of the integrand only contains $q$ in the term
$\vec{\sigma}\cdot\vec{q}\ \vec{\sigma}\cdot\vec{p}$. This term vanishes 
exactly in the limit $\vec{p_1}\to 0$, or in any case if one makes the 
heavy baryon approximation 
${p_1}^0-E(\vec{p_1}+\vec{p})=0$. 
(The corrections are of order ${(\vec{p_1}/M)}^2$, which we can neglect.) 
We shall evaluate the $NN$ potential for $\vec{p_1}=0$ for simplicity.
With the $\vec{\sigma}\cdot\vec{q}$ $\vec{\sigma}\cdot\vec{p}$ term of Eq. (\ref{E12}) 
vanishing, we observe that $V^{(A)}$ and $V^{(B)}$ are equal but with opposite 
sign, and hence there is an exact cancellation of these two terms.

 If now we take the off shell part of the meson vertex corresponding to the 
other mesons we would observe an exact cancellation with the diagrams 
$(b)$-$(d)$ of Fig. \ref{F2}. This is interesting for practical purposes 
since it means at the end that we must evaluate only the diagrams of Fig. \ref{F1} and using only the on-shell value of the meson-meson vertex.

 The need to include this subset of chiral diagrams to find cancellations in 
the $NN$ isoscalar interaction was already stressed in Ref. [9]. There, 
different arguments were used, recalling that in the expansion of the 
pion field matrix $U(\vec{\pi})$, the third order term has ambiguities and 
the result cannot depend upon them.
The on-shell amplitude does not depend on these ambiguities, 
so here we also prove that, 
after summing the terms discussed above, the results do not depend on unknown 
parameters which would affect the off shell meson-meson amplitude.

\begin{figure}[ht]
\begin{center}
\epsfxsize=14cm
\centerline{\epsfbox{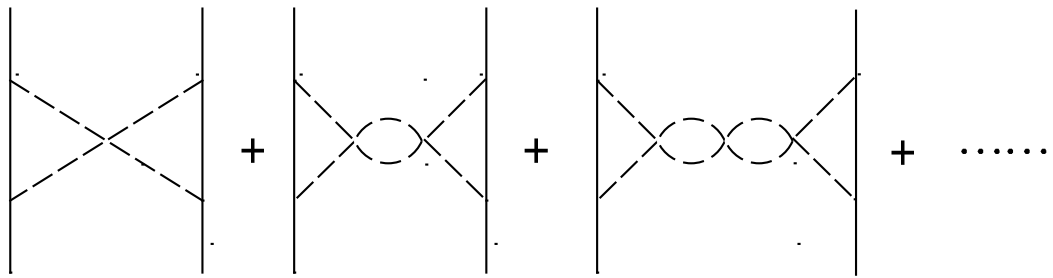}}
%\epsfile{file=figure5.eps,scale=0.8}
\caption{The nucleon-nucleon interaction in the scalar-isoscalar channel, where the $\pi\pi$ scattering $t$-matrix is summed up to all orders in the unitary approach.}
\label{F5}
\end{center}
\end{figure}

 The derivation here is particularly useful for our purposes, since in Ref. 16) 
it was also shown that in the construction of the full meson-meson 
amplitude, only 
the on-shell part of the meson-meson vertex was needed. This allows us to sum 
immediately the set of diagrams that were included in the unitary Bethe-Salpeter 
approach to the scalar meson amplitude and, hence, in the $NN$ 
interaction we would have the set of diagrams shown in Fig. \ref{F5}.

\newpage

\section{Unitary approach for the interacting pions in the isoscalar chnnel}

 The diagrams of Fig. \ref{F5} are easily summed up. First one must 
substitute the on-shell lowest-order meson-meson amplitude of Eq. (\ref{E6}) 
with the Bethe-Salpeter amplitude,~\cite{R16}
\begin{equation}
t^{(I=0)}_{\pi\pi\to\pi\pi}=-\frac{1}{f^2}\frac{(s-\frac{{m_\pi}^2}{2})}
{1+\frac{1}{f^2}(s-\frac{{m_\pi}^2}{2})G(s)}.
\label{E14}
\end{equation}
The function $G(s)$ is the loop function with two pion propagators, 

\begin{equation}
G(s)=i{\int}\frac{d^4q}{{(2\pi)}^4}\frac{1}{q^2-{m_{\pi}}^2+i\epsilon}\ 
\frac{1}{{(P-q)}^2-{m_{\pi}}^2+i\epsilon},
\label{E15}
\end{equation}
where $P$ is the total momentum of the two-pion system and $P^2=s$.
In Ref. [16] the integral in Eq. (\ref{E15}) was regularized with a cut-off in 
the $CM$ frame of the two mesons.
Here it is better to work in the $CM$ of the two nucleons, and hence 
an invariant form for $G(s)$ is preferable.
This can be done by using the results for $G(s)$ obtained with dimensional 
regularization, which are shown in Appendix A of Ref. [20] to be equivalent 
to those of a cut-off. We then have
\begin{equation}
G(s)=\frac{1}{{(4\pi)}^2}\left[-1+\ln\frac{{m_{\pi}}^2}{\mu^2}+
{\sigma}\ln\frac{\sigma+1}{\sigma-1}\right],
\label{E16}
\end{equation}
with $\mu$ the regularization mass, which was found to be $\mu=1.2q_
{\rm max}=1.1$ GeV 
for the value of the cutoff, $q_{\rm max}$, needed for a good fit to the data with the B-S equation and the lowest order 
${\chi}PT$ as the kernel of the equation.
In Eq. (\ref{E16}) the magnitude of $\sigma$ is 

\begin{eqnarray}
\sigma=\sqrt{1-\frac{4{m_{\pi}}^2}{s}},
\end{eqnarray}
In the range $0<s<4{m_{\pi}}^2$, one has an analytical extrapolation which 
leads to 

\begin{equation}
{\sigma}\ln\frac{\sigma+1}{\sigma-1}\to(\pi-2\alpha)\sqrt{\frac{4{m_{\pi}}^2}
{s}-1};\ \alpha=\arctan\sqrt{\frac{4{m_{\pi}}^2}{s}-1},
\label{E17}
\end{equation}
and for $s>4{m_{\pi}}^2$ the log of Eq. (\ref{E16}) develops a negative 
imaginary part. For $s<0$, $\sigma$ is always larger than 1, and the log term behaves 
smoothly.

\begin{figure}[h]
\begin{center}
\epsfxsize=4cm
\centerline{\epsfbox{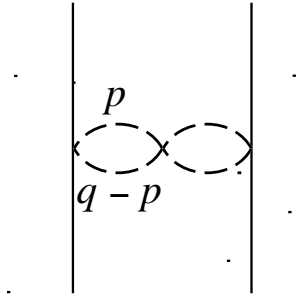}}
%\epsfile{file=figure6.eps,scale=0.8}
\caption{The nucleon-nucleon interaction with the two-pion nucleon vertex in the isoscalar channel.}
\label{F6}
\end{center}
\end{figure}

 We would also like to note that at the same order as calculated in this 
section there are other pieces, like in Fig. \ref{F6}. The $BBMM$ contact term 
is proportional to $\gamma^\mu \tau^a$, which implies vector and iso-vector 
exchange in the $t$ channel, and hence we do not have
to consider it for the isoscalar exchange channel we are concerned with here.

 In the $NN$ $CM$ frame we have the momentum $q$ exchanged between the nucleons
 with $q\equiv(0,\vec{q})$, such that $s=-{\vec{q}}^{\ 2}$. The $NN$ potential 
corresponding to the isoscalar $2\pi$ exchange of Fig. \ref{F5} then becomes
\begin{equation}
t_{NN}(q)={{\widetilde{V}}_N}^2(q)\frac{6}{f^2}\frac{({\vec{q}}^{\ 2}+\frac{
{m_{\pi}}^2}{2})}{1-G(-{\vec{q}}^{\ 2})\frac{1}{f^2}({\vec{q}}^{\ 2}+
\frac{{m_{\pi}}^2}{2})}.
\label{E18}
\end{equation}
Here $\widetilde{V}(q)$ is the vertex corresponding to the triangle loop to the 
left in Fig. \ref{F4}, which in the limit $\vec{p_1}\to0$ can be written as

\begin{eqnarray}
{\widetilde{V}}_N(q)&=&i\int\frac{d^4p}{{(2\pi)}^4}{\left(\frac{D+F}
{2f}\right)}^2
{\vec{\sigma}}\cdot(\vec{p}+\vec{q})\vec{\sigma}\cdot\vec{p}\ 
\frac{1}{p^2-{m_{\pi}}^2+i\epsilon}\ \frac{1}{(p+q)^2-{m_{\pi}}^2+i\epsilon}
\nonumber \\
&&\frac{M}{E(\vec{p})}\ \frac{1}{M+p^0-E(\vec{p})+i\epsilon}.
\label{E19}
\end{eqnarray}
With the sums over the spins of the intermediate nucleon in the loop we have

\begin{equation}
\vec{\sigma}\cdot(\vec{p}+\vec{q})\vec{\sigma}\cdot\vec{p}=(\vec{p}+\vec{q})
\cdot\vec{p}+(\vec{q}\times\vec{p})\cdot\vec{\sigma},
\label{E20}
\end{equation}
but the integral over $\vec{p}$ in Eq. (\ref{E19}) leads to a vector 
proportional to $\vec{q}$, which makes the term with $\vec{\sigma}$ of 
Eq. (\ref{E20}) vanish in the integral.
Furthermore, the $p^0$ integration in Eq. (\ref{E19}) can be done 
analytically, and one finds

\begin{eqnarray}
\widetilde{V}_N(q)&=&{\int}\frac{d^3p}{(2\pi)^3}\left(\frac{D+F}{2f}\right)^2
(\vec{p}^{\ 2}+\vec{p}\cdot\vec{q})\frac{M}{E}\ \frac{1}{2}\ \frac{1}{\omega}
\ \frac{1}{\omega'}\ \frac{1}{\omega+\omega'}
\nonumber \\
&&\frac{1}{E+\omega-M}\ \frac{1}{E+\omega'-M}
[\omega+\omega'+E-M],
\label{E21}
\end{eqnarray}
with

\begin{equation}
E=E(\vec{p});\ \omega=\sqrt{{m_{\pi}}^2+\vec{p}^{\ 2}};\ 
\omega'=\sqrt{{m_{\pi}}^2+(\vec{p}+\vec{q})^{\ 2}},
\label{E22}
\end{equation}
The integral in Eq. (\ref{E21}) is logarithmically divergent.

\begin{figure}
\begin{center}
\epsfxsize=4cm
\centerline{\epsfbox{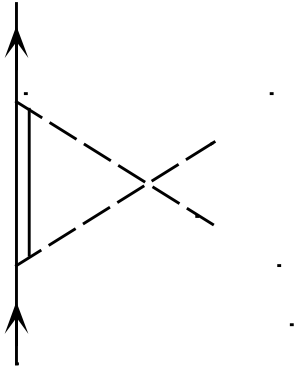}}
%\epsfile{file=figure7.eps,scale=0.8}
\caption{The two pion exchange triangle vertex through delta excitation.}
\label{F7}
\end{center}
\end{figure}

It is straighforward to include the contribution with an intermediate 
$\Delta$ instead of a nucleon in the vertex function, as depicted in 
Fig. \ref{F7}.
It can be evaluated in a straighforward way, substituting $S_iT_j$ for 
$\sigma_i\tau_j$ with $S$ and $T$ the spin and isospin transition 
operators.

 The inclusion of the $\Delta$ in the fermionic loops is required in order 
to satisfy the basic rules in the large $N_c$ limit,~\cite{R21} and we include it here. We obtain

\begin{eqnarray}
\widetilde{V}_{\Delta}(q)&=&\frac{4}{9}{\left(\frac{f^*_{{\pi}N{\Delta}}}
{f_{{\pi}NN}}\right)}^2
\int\frac{d^3p}{(2\pi)^3}\left(\frac{D+F}{2f}\right)^2
(\vec{p}^{\ 2}+\vec{p}\cdot\vec{q})\frac{M_\Delta}{E_\Delta}
\frac{1}{2}\frac{1}{\omega}\frac{1}{\omega'}
\nonumber \\
&&\frac{1}{\omega+\omega'}\ 
\frac{1}{E_\Delta+\omega-M}\ \frac{1}{E_\Delta+\omega'-M}
[\omega+\omega'+E_\Delta-M],
\label{E23}
\end{eqnarray}
with $M_\Delta$ the $\Delta$ mass and $E_\Delta=({M_\Delta}^2+
\vec{p}^{\ 2})^{1/2}$. We take the empirical value for the ratio of the
couplings $f^*_{\pi N\Delta}$ to $f_{\pi NN}$ as 2.12 .

 The integral of Eq. (\ref{E23}) is also logarithmically divergent.
In order to regularize the integrals of Eqs. (\ref{E21}) and (\ref{E23}), 
one needs a cutoff or a form factor.
Within a quark model one finds natural form factors due to the finite size of 
the nucleon, but one must still sum over excited states of the quarks 
in the intermediate states, which leads to a divergence in spite of the 
form factors.~\cite{R22} The regularization of the loop in this case can be 
accomplished by a cutoff in the space of intermediate states, together 
with a form factor. The truncation at $\Delta$, together with 
the use of monopole form factors with $\Lambda$ of order 1 GeV 
has been found to be a sensible regularization procedure, as proved by the success 
of such an approach in the cloudy bag model.~\cite{R23}
We shall use here static form factors to keep the $p^0$ integral of 
Eq. (\ref{E19}) simple, and hence we include in the integrand of Eqs. (\ref{E21})
 and (\ref{E23}) the product of form factors

\begin{equation}
F(\vec{p})F(\vec{p}+\vec{q})=\frac{\Lambda^2}{\Lambda^2+\vec{p}^{\ 2}}
\ \frac{\Lambda^2}{\Lambda^2+(\vec{p}+\vec{q})^{\ 2}},
\label{E24}
\end{equation}
with $\Lambda\approx1-1.2$ GeV.~\cite{R1}

 Our final form for the $NN$ potential in momentum space is thus given by

\begin{equation}
t_{NN}(q)={\widetilde{V}(q)}^2\frac{6}{f^2}\ 
\frac{(\vec{q}^{\ 2}+\frac{{m_{\pi}}^2}{2})}{1-G(-\vec{q}^{\ 2})\frac{1}{f^2}
(\vec{q}^{\ 2}+\frac{{m_{\pi}}^2}{2})},
\label{E25}
\end{equation}
with

\begin{equation}
\widetilde{V}(q)=\widetilde{V}_N(q)+\widetilde{V}_\Delta(q).
\label{E26}
\end{equation}
Here $\widetilde{V}_N(q)$ and $\widetilde{V}_\Delta(q)$ are given by Eqs. (\ref{E21}) and 
(\ref{E23}) incorporating in the integrand the product 
$F(\vec{p})F(\vec{p}+\vec{q})$ 
of Eq. (\ref{E24}).
The potential in coordinate space is given by 
\begin{eqnarray}
V^{(S)}_{NN}(r)&=&\int\frac{d^3q}{(2\pi)^3}e^{i\vec{q}\vec{r}}t_{NN}(q)
\nonumber \\
&=&\frac{1}{2\pi^2}\frac{1}{r}\int_0^{\infty}q{}dq{}\sin(qr)t_{NN}(q).
\label{E27}
\end{eqnarray}

We would like to compare $V^{(S)}_{NN}(r)$ with the empirical $\sigma$ exchange given by

\begin{eqnarray}
V^{(\sigma)}_{NN}(r)&=&\int\frac{d^3q}{(2\pi)^3}e^{i\vec{q}\vec{r}}
\frac{g^2_{{\sigma}NN}}{-\vec{q}^{\ 2}-{m_{\sigma}}^2}
\nonumber \\
&=&-\frac{g^2_{{\sigma}NN}}{4\pi}\frac{e^{-m_{\sigma}r}}{r},
\label{E28}
\end{eqnarray}
with

\begin{equation}
\frac{{g_{{\sigma}NN}}^2}{4\pi}=5.69;\ g_{{\sigma}NN}=8.46,
\label{E29}
\end{equation}
from Ref. [1]. This corresponds to the empirical $\sigma$ exchange when the two pion box diagrams with the intermediate $\Delta$ are considered in addition. 

Actually, in Ref. [1] a monopole form factor per vertex with $\Lambda_
{\sigma}=1.7$ GeV is also included so that

\begin{eqnarray}
V_{NN}^{(\sigma)}(q)=-g_{{\sigma}NN}^2
\left(\frac{\Lambda^2_{\sigma}-m^2_{\sigma}}{\Lambda^2_{\sigma}+\vec{q}^{\ 2}}
\right)^2
\frac{1}{m_{\sigma}^2+\vec{q}^{\ 2}},
\end{eqnarray}
which leads to the potential in coordinate space

\begin{eqnarray}
V_{NN}^{(\sigma)}(r)=\frac{1}{4\pi}g_{{\sigma}NN}^2
\left\{\frac{1}{r}e^{-\Lambda_{\sigma}r}+\frac{\Lambda^2_{\sigma}-
m^2_{\sigma}}{2\Lambda_{\sigma}}e^{-\Lambda_{\sigma}r}
-\frac{1}{r}e^{-m_{\sigma}r}
\right\},
\label{E50}
\end{eqnarray}
with the same asymptotic behaviour as Eq. (\ref{E28}) at large distances and 
the finite limit $-(\Lambda_{\sigma}-m_{\sigma})^2/2\Lambda_{\sigma}$ for 
$r\to0$.

\section{Qualitative discussion of the potential}

 Let us go back to Eq. (\ref{E14}) in the physical region.
We know from Ref. [16] that the isoscalar amplitude develops a pole around 
$450-i225$ MeV. Let us neglect for this study Im$G(s)$ and rewrite 
Eq. (\ref{E14}) as

\begin{equation}
t^{(I=0)}_{\pi\pi\to\pi\pi}=
\frac{-G(s)^{-1}(s-\frac{{m_{\pi}}^2}{2})}{s-(\frac{{m_{\pi}}^2}{2}
-f^2G(s)^{-1})},
\label{E31}
\end{equation}
so that

\begin{equation}
\frac{{m_{\pi}}^2}{2}-f^2G(s)^{-1}\simeq{m_\sigma}^2;\ 
G(s)^{-1}=\frac{1}{f^2}(\frac{{m_{\pi}}^2}{2}-{m_\sigma}^2).
\label{E32}
\end{equation}
Assuming now that $G(s)^{-1}$ is a smooth function of $s$ and $\widetilde{V}(q)$ 
is a smooth function of $q$, we would have

\begin{eqnarray}
t_{NN}(q)&\simeq&{\widetilde{V}(0)}^2\frac{6}{f^2}
\frac{({m_{\sigma}}^2-\frac{{m_{\pi}}^2}{2})(\vec{q}^{\ 2}+
\frac{{m_{\pi}}^2}{2})}{\vec{q}^{\ 2}+{m_{\sigma}}^2}
\nonumber \\
&=&{\widetilde{V}(0)}^2\ \frac{6({m_{\sigma}}^2-\frac{{m_{\pi}}^2}{2})}{f^2}
\ \frac{(\vec{q}^{\ 2}+{m_{\sigma}}^2)-({m_{\sigma}}^2-\frac{{m_{\pi}}^2}{2})}
{\vec{q}^{\ 2}+{m_{\sigma}}^2}.
\label{E33}
\end{eqnarray}
We see that the first term in the numerator of the second fraction 
gives rise to a repulsive $\delta$ function in coordinate space,
while the second term gives rise to an attractive potential due to a standard 
$\sigma$ exchange with the equivalent coupling

\begin{equation}
g_{{\sigma}NN}\simeq\widetilde{V}(0)\sqrt{6}
\frac{({m_\sigma}^2-\frac{{m_{\pi}}^2}{2})}{f}
\label{E34}
\end{equation}
With the value $\widetilde{V}(0)\simeq 0.10\times 10^{-2}$ MeV$^{-1}$ 
that we obtain, as we shall see, the $\sigma NN$ coupling would 
be of the order 
of $g_{{\sigma}NN}\simeq 5$, which has the right order of magnitude, 
compared with the empirical coupling used in Ref. [1] to provide the needed 
intermediate range attraction, $g_{{\sigma}NN}\simeq 8.5$.

 This qualitative study is certainly improved by the accurate results which 
we give in the next section, but it serves to illustrate the features 
which we can expect from such a potential. The interesting thing is that, 
apart from the intermediate attraction, one gets a scalar repulsion, 
which is also demanded by the $NN$ scattering data.

\section{Results and discussion}

   The qualitative discussion of \S 4 relies on the constancy of
the $G(s)$ function of Eq. (\ref{E16}). This function is rather smooth and
is negative in the range of momenta of interest. For $s$ = 450 
MeV$^2$
it is about $-0.02$, and in the range of $s$ from 0 to $-500$ MeV$^2$ it varies
between $-0.02$ and $-0.015$ .
The function $\widetilde{V}(q)$ changes slightly faster and drops by a factor
of 2 as $q$ varies from 0 to 550 MeV. Hence the qualitative results of that section are
only indicative of the actual numerical results.  We performed the
calculations using  a value of $\Lambda$ for the monopole form factors of
Eq. (\ref{E24}) of 1.1 GeV.

\begin{figure}[h]
\begin{center}
\epsfxsize=13cm
\centerline{\epsfbox{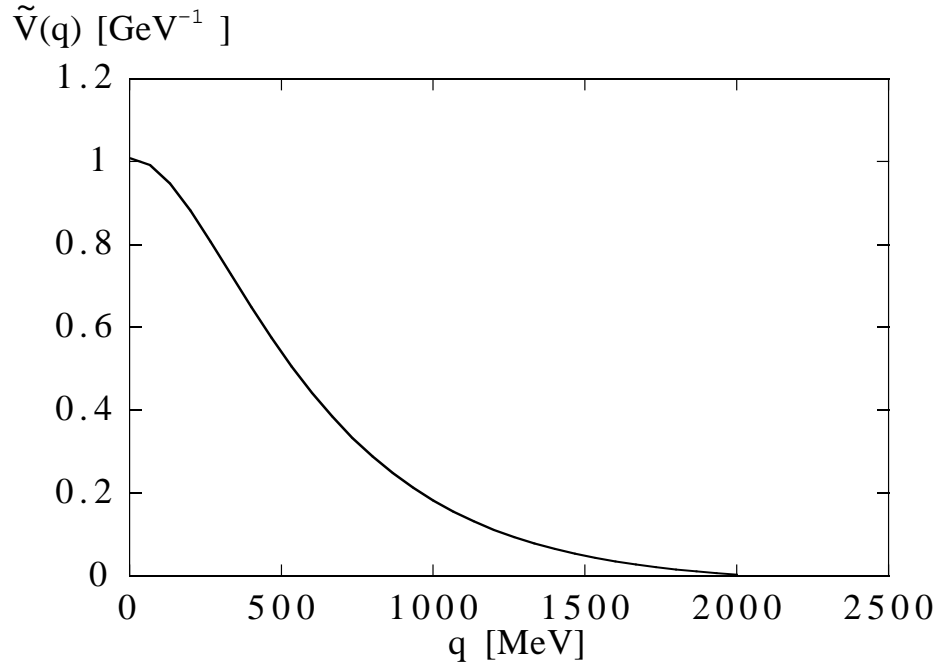}}
%\epsfile{file=figure8.eps,scale=0.5}
\caption{The vertex function $\widetilde{V}(q)$ as a function of momentum $q$.}
\label{F8}
\end{center}
\end{figure}

  In Fig. \ref{F8} we give the results for the vertex function, which display a
moderate but steady decrease of the vertex with $q$. This ensures
convergence of the integral of Eq. (\ref{E27}) for $V(r)$.  At this point we
note already some discrepency with Ref. [9], where no
form factors were used in the evaluation of the vertex function. The
divergences obtained there led to singularities at short distances which
were disregarded, since only the peripheral $NN$ partial waves were
investigated there. In the present case, both the $q$ dependence of the
form factor and the extra denominator in Eq. (\ref{E25}) from the
iteration of unitary loops, which produces the sigma resonance, improve
the convergence properties of the integral of Eq. (\ref{E27}).
\begin{figure}[h]
\begin{center}
\epsfxsize=13cm
\centerline{\epsfbox{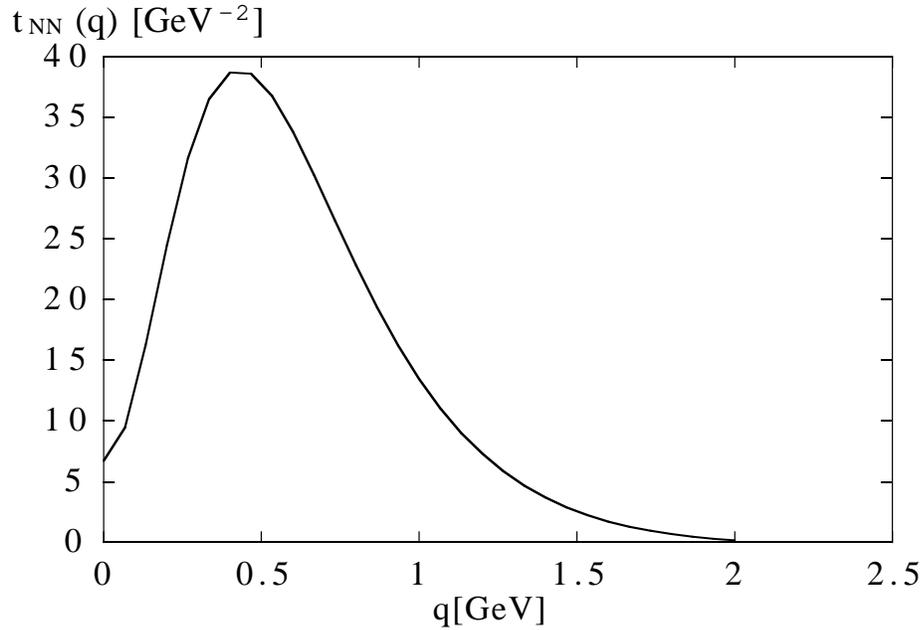}}
%\epsfile{file=figure9.eps,scale=0.5}
\caption{The function $t_{NN}$.}
\label{F9}
\end{center}
\end{figure}
The function $t_{NN}(q)$ of Eq. (\ref{E25}) is plotted in Fig. \ref{F9}. It exhibits an
increase from 0 to about 500 MeV and then drops as q increases further. For $q$
around 2 GeV, it has a value about 250 times smaller than at the maximum.

We should caution that the unitary theory used here leads to good
results in the physical region up to about $\sqrt{s}$ = 1.2 GeV.  It is then 
clear that one should not extrapolate the results of the model in the
unphysical region to values of $q$  much larger than 1 GeV. In this
sense, the form factors in the ${\pi}NN$ vertices with lambda of order
 1 GeV guarantee that one does not enter into this unknown region. In
any case, the uncertainties  for these large values of $q$ would revert
into uncertainties at  short distances in $V(r)$, which we certainly must
admit. Technically, the function $t_{NN}(q)$ of Eq. (\ref{E25}) develops a pole
around 2060 MeV, far away from the region of validity of the model used, 
and to which we do not give any physical meaning. This is anyway a
warning that we should not attempt to investigate the behaviour of the
potential at very short distances with the present model. The numerical
integrations are extended up to 2 GeV in the actual calculations. The
predictions for $V(r)$ beyond $r$ = 0.5 fm should be rather safe.

\begin{figure}
\begin{center}
\epsfxsize=13cm
\centerline{\epsfbox{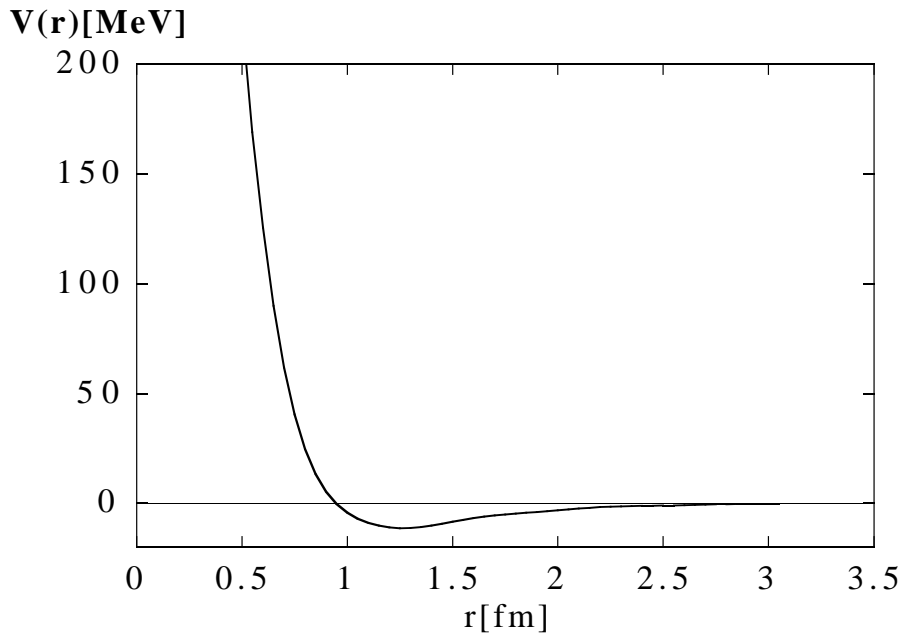}}
%\epsfile{file=figure10.eps,scale=0.5}
\caption{The $NN$ potential in coordinate space}
\label{F10}
\end{center}
\end{figure}

   In Fig. \ref{F10} we show the results for $V(r)$. We find interesting
behaviour with a moderate attraction beyond $r$ = 0.9 fm and a repulsion at
shorter distances, which become of order 1 GeV at very short
distances, with the caveat discussed above.  Around 0.5 fm it has a
value of around 200 MeV. The trend of the results agrees with the
qualitative results anticipated in \S 4.
  The first thing to note is that these results are quite different
qualitatively from those of the conventional sigma exchange, where there is
only attraction, according to Eq. (\ref{E50}).  They also differ appreciably
from those of Ref. [9] where a very weak repulsion is obtained in the region
where we find an attraction here (the attraction of the order of eV
found numerically at larger distances in Ref.9 is not significative). This
finding is relevant since it shows that even for large distances, which
is the focus of Ref. [9], one needs to go beyond perturbation theory in the $\pi \pi$ interaction in order to determine the behaviour of the potential.
    A study of the behaviour of $V(r)$ around 2.5 fm reveals Yukawa
behaviour with $m_\sigma$ around 450 MeV and a coupling $g_{\sigma NN}$ 
of about
8, in rough agreement with the results of the qualitative discussion of
\S 4, demonstrating that the sigma pole appearing in the physical region
shows up indeed in the potential at large distances.  Yet, the presence
of the sigma pole is not the only element responsible for the behaviour
of the potential seen in Fig. \ref{F10}. Indeed, if we replace the denominator
of Eq. (\ref{E25}) by unity, thus removing the sigma pole, we still find
structure of the potential similar to that in Fig. \ref{F10}, with the
attraction appearing at shorter distances beyond $r$ = 0.85 fm. The strength
of the attraction is larger and also falls off faster with increasing $r$,
indicating that the range is now given by other $q$-dependent functions.
In this case, the vertex function and its $q$ dependence are responsible for
that behaviour and for the differences with the results of Ref. [9]. Both the vertex
function and the unitary approach to the $\pi \pi$ interaction are thus very
important for the isoscalar potential and for determining its strength
in all ranges of distances. Note that the phenomenological form factor 
that we assume modulates the structure of the vertex.  It not only affects the 
vertex at large $q$, in which
case only the short range part of the interaction would be modified, but it
also modifies the strength of the vertex at finite $q$, and this also has
influence in the medium and long range parts of the interaction.
For similar reasons, the unitarization changes the $q$ dependence, but
more importantly it introduces a pole in the $s$ channel, which is reflected
in the t channel by a distinct tail in $r$ that goes roughly as
$e^{-m_{\sigma}r}/r$.

  The evaluation of the vertex function here has been done using
elements of phenomenology, beyond the chiral approach otherwise used.
In  this sense the monopole form factors were used and the space of
intermediate baryon states was truncated in the delta. Other more
realistic options could be used, maybe even using form factors obtained
within the chiral approach in some self-consistent way. For the time
being we should admit uncertainties from this source.These uncertainties should not
be minimized since the results depend appreciably on the choice of the
range parameters. For instance a change of lambda from 1.1 GeV to 1.2
GeV changes the attraction from 10 to 15 MeV. A different vertex
evaluation, in which an explicit contribution of states beyond the delta
in the intermediate states is allowed, as in Ref. [22] would also lead to
different $q$ behaviour of the vertex function, with immediate results in
the potential in coordinate space. The regularization procedure can be done 
in many ways.  In chiral perturbation, it requires a regularization scheme 
by means of a regularization scale or a cut-off, but at the same time one 
has to introduce counter-terms of higher order.  In the case of the 
meson-meson interactions, it is proved in Ref. [17] and [20] that a cut-off, or 
equivalently dimensional regularization at a certain scale, effectively 
generates the counter-terms in the meson-meson $s$-wave channel.  For the 
vertex functions, which we generate here, such a procedure is not yet 
available. In Ref. [8] and [9], infinities in the vertex function
are removed with the claim that this removal affects only the short range
part of the interaction.  In our case we have resorted to phenomenology 
using an empirical $\pi NN$ form factor  which makes the loop
function convergent.  The fact that with this procedure we find an
assymptotic behavior for the potential similar to the empirical 
$\sigma$ potential gives us confidence that this ``empirical'' regularization 
in the baryon sector is a reasonable procedure.

In any case, given all the limitations which have been discussed above, we can
not make strong claims about the strength of this $\sigma$ potential,
but we certainly can stress the qualitative features, which are rather
different from the conventional results for sigma exchange and from
results obtained in the perturbative approach to the problem.

\section{Conclusions}

We have evaluated the contribution to the $NN$ potential from the exchange
of an interacting pion pair in the scalar-isoscalar channel, the sigma
channel. We have used  a unitary model based upon the lowest order
chiral Lagrangian and the Bethe-Salpeter equation. This model is supported
by a more elaborate approach using the inverse amplitude method and the
lowest and second order chiral Lagrangians, but technically simpler.
Both approaches generate a sigma pole around 450 MeV, although with a
large width. The $\pi \pi$ amplitude extrapolated to negative values of $s$,
retains memory of the sigma pole and leads to a tail of the $NN$ potential
reflecting a Yukawa behaviour with the range of the sigma mass.  We
obtain some moderate attraction beyond 0.9 fm and a stronger repulsion
before that. This behaviour is quite different from that of the
conventional sigma exchange, which always leads to an attractive force,
even in the presence of form factors for the $\sigma NN$ vertices.

   We have also found that the procedure chosen to regularize the $\sigma NN$
vertex function also has important consequences for the potential.  We
have chosen to introduce phenomenological form factors for the $\pi NN$
vertices and to restrict the space of baryon intermediate states in the
loop to the nucleon and the delta, a procedure which has been often used
in related problems in connection with chiral quark models. The results
are sensitive to the form factors and the manner in which the vertex is constructed.
This is one of the problems that requires further thought. For the
moment, we should accept one degree of freedom due to the
divergence of the vertex loop, which must be cured with some
regularization mass, a counterterm, the use of a form factor, etc. We
can hope to reduce the freedom by using some empirical information, as
in the present case . Hence we believe our results are qualitatively
acceptable in the range of distance beyond 0.5 fm, and more quantitatively
accurate results. The findings of
this paper are important because they reveal behaviour
that is quite different from that which one gets from the conventional sigma exchange picture.
On the other hand, it is also known that a nonperturbative approach to the
$\pi \pi$ interaction is needed, and such an approach should be important for 
obtaining proper results in the whole range of 
distances in the $NN$ interaction. These findings should thus have
important consequences in the chiral approach to the $NN$ interaction, 
which is now the subject of much theoretical investigation.

\section*{Acknowledgements}
We are grateful to the COE Professorship program of Monbusho, which
enabled E. O. to stay at RCNP to perform the present work. One of us,
E. O., wishes to thank N. Kaiser for useful discussions. This work is
partly supported by DGICYT contract number PB96-0753.

\end{document}